\documentclass[twocolumn, floatfix, prb, superscriptaddress, aps,showpacs,longbibliography]{revtex4-2}
\usepackage{graphicx,amsmath,amssymb, color, bm,mathtools}
\usepackage{mathrsfs}
\usepackage{nicefrac}
\usepackage{fancyhdr, lipsum}
\usepackage[titletoc,title]{appendix}
\usepackage[dvipsnames]{xcolor}
\usepackage[normalem]{ulem}
\usepackage{orcidlink}
\usepackage{dcolumn}
\usepackage{physics}

\newcommand{\be}{\begin{equation}}
\newcommand{\ee}{\end{equation}}

\newcommand{\ba}{\begin{eqnarray}}
\newcommand{\ea}{\end{eqnarray}}

\newcommand{\LLs}{$\Lambda$Ls }

\newcommand{\LLL}{\mathcal{P}_\mathrm{LLL}}

\begin{document}

\title{Universality of long-wavelength behavior of composite-fermion Fermi liquid}

\author{Aamir A. Makki}
\email{makki@psu.edu}
\affiliation{Department of Physics, 104 Davey Lab, Pennsylvania State University, University Park, Pennsylvania 16802, USA}
\author{Mytraya Gattu}
\email{mvg6042@psu.edu}
\affiliation{Department of Physics, 104 Davey Lab, Pennsylvania State University, University Park, Pennsylvania 16802, USA}
\author{J. K. Jain\orcidlink{000-0003-0082-5881}}
\affiliation{Department of Physics, 104 Davey Lab, Pennsylvania State University, University Park, Pennsylvania 16802, USA}
\affiliation{Center for Theory of Emergent Quantum Matter, Pennsylvania State University, University Park, Pennsylvania 16802, USA}
\affiliation{Lodha Theoretical Physics Institute, 17th Floor, Lodha NCP Supremus, Wadala (E), Mumbai 400037, India}
\email{jkj2@psu.edu}

\begin{abstract}
A recent article evaluated the long-wavelength behavior of the projected static structure factor of the composite-fermion (CF) liquid within the zeroth-order microscopic theory and found $\bar{S}(\mathbf{q})\sim q^3$, in disagreement with the $\bar{S}(\mathbf{q})\sim q^3\ln q$ behavior predicted by the Chern-Simons field theory for the Coulomb interaction. Here we consider the possibility that the discrepancy arises because the zeroth-order CF Fermi-liquid wave function used in that work does not properly capture the long wavelength behavior.  We use CF diagonalization to significantly improve the wave function but do not find any evidence for $\bar{S}(\mathbf{q})\sim q^3 \ln q$ behavior. Additionally, we find that the small-$q$ behavior of $\bar{S}(\mathbf{q})$ is also insensitive to the form of the interaction between electrons, suggesting universality.
\end{abstract}

\maketitle

{\it Introduction:} The central principle of the composite-fermion (CF) theory is that electrons confined to two dimensions and exposed to a high magnetic field bind an even number of quantized vortices to form composite fermions, which experience a reduced effective magnetic field~\cite{Jain89,Halperin93,Jain07,Jain20,Halperin20,Stormer07,Shayegan20}. It provides a remarkably successful and unified microscopic account of a vast zoo of strongly-correlated electronic states in terms of weakly interacting CFs. In particular, it predicts fractional quantum Hall (FQH) effects at the Jain fractions arising from the integer quantum Hall (IQH) states of CFs, and metallic, or Fermi liquid, states of CFs at even-denominator filling factors\cite{Jain07,Jain20}. Many hallmarks of the CF Fermi liquids have been confirmed experimentally~\cite{Willett93,Goldman94,Kang93,Du94,Hossain20,Kamburov14b,Willett99,Shayegan20}. This article concerns a subtle long-distance property of these states. 

An important quantity characterizing these Fermi liquids is the projected static structure factor $\bar{S}(\mathbf{q})$, defined below, which probes density correlations within the lowest Landau level (LLL)~\cite{Girvin85,Girvin86,He94,Dora24}. In particular, the long-wavelength behavior of $\bar{S}(\mathbf{q})$ provides a test of competing theoretical descriptions. A recently developed method~\cite{Gattu25} has made it possible to study much larger systems than before using the microscopic wave functions of the CF theory, and a calculation using the CF Fermi sea wavefunction found that $\bar{S}(\mathbf{q})$ behaves as $q^3$ at small $q$ \cite{Anakru25}, in contrast with the Chern-Simons/random phase approximation (RPA) field-theoretic treatments, first performed by Halperin, Lee and Read (HLR), which predict a $q^3\ln q$ form for the Coulomb interaction\cite{Lopez91,Halperin93,Simon93,Simon94a,Son15,Read98,Simon98,Halperin03,Hofmann21,Murthy02}. Earlier studies including CFs in a spherical geometry and DMRG on a cylinder addressed the same question but did not definitively resolve the thermodynamic small-q behavior\cite{Balram17,Kumar22}. More generally, Fermi surfaces coupled to emergent gauge fields are expected to exhibit singular non-Fermi-liquid corrections beyond ordinary Landau Fermi-liquid theory~\cite{Polchinski94,sung-sik-lee-2009,lee_nagaosa_wen2006,senthil2008,sachdevQPT,Sachdev10,Lee18,Varma02}. A complementary viewpoint is provided by the dipole picture of the composite fermion, in which the composite fermion is viewed as an electron bound to correlation holes/vortices and behaves as a neutral dipole\cite{Pasquier98,Predin2023DipoleHalf,Murthy03}. The dipole picture was found to give a $\bar S(\mathbf{q})\sim q^3$ long-wavelength behavior for the CF Fermi sea, in agreement with the zeroth-order CF Fermi sea wavefunction\cite{Anakru25}.

A possible resolution of the discrepancy is that the zeroth-order CF liquid wave function does not capture the long distance behavior of the true CF liquid.
The purpose of the present work is to address this possibility. While the zeroth-order CF theory already provides an extremely accurate representation of the ground state of interacting electrons, it can be further improved in a systematic manner by the method called CF diagonalization (CFD)~\cite{Mandal01a,Mandal02,Jain07,Scarola00,Balram13}. We demonstrate below that using this method we are able to obtain a significantly improved wave function for the CF Fermi liquid, which, our results strongly suggest, is close to the exact state. However, we find that $\bar{S}(\mathbf{q})$ remains essentially unaltered.

\begin{figure*}
    \centering
    \includegraphics[width=0.99\linewidth]{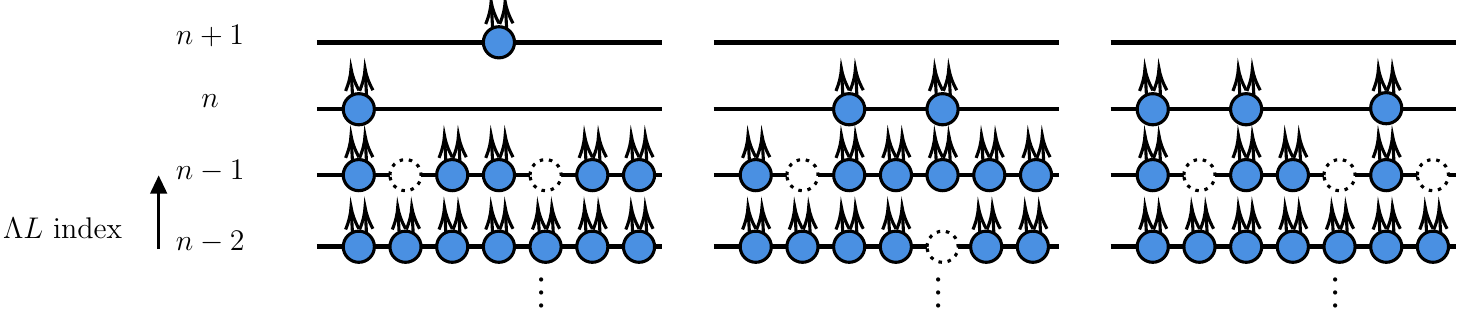}
    \caption{Examples of basis functions with CFKE$=3$. A single CF exciton with CFKE$=3$ is not included as it has $L\geq 3$ and thus does not mix with the $L=0$ ground state.
   }
    \label{fig:3ke_exciton}
\end{figure*}

\begin{table*} \caption{Ground-state energy $E_0^{\alpha}$ obtained by CFD. The energies are quoted in units of $e^2/(\epsilon\ell_B)$, where $\epsilon$ is the dielectric constant of the background semiconductor, and only represent the electron-electron interaction component; electron-background and background-background interaction energies are not included. For each $\alpha$ and $N$, we also give the dimension of the CF basis $D^{\alpha}_{\rm CF}$, that is, the number of linearly independent $L=0$ states in the basis in which the Hamiltonian has been diagonalized to obtain $E_0^{\alpha}$; the number of states eliminated in going from the basis in Eq.~\ref{eq:basis_electron} to that in Eq.~\ref{eq:basis_CF} is given in parentheses. } \label{tab:ke_table_GSE} 
\centering \begin{ruledtabular} \begin{tabular}{c D{.}{.}{4} D{.}{.}{4} D{.}{.}{4} c c c} \multicolumn{1}{c}{$N$} & \multicolumn{1}{c}{$E_0^{0}$} & \multicolumn{1}{c}{$E_0^{2}$} & \multicolumn{1}{c}{$E_0^{3}$} & \multicolumn{1}{c}{$D^{\alpha=0}_{\rm CF}$} & \multicolumn{1}{c}{$D^{\alpha=2}_{\rm CF}$} & \multicolumn{1}{c}{$D^{\alpha=3}_{\rm CF}$} \\ \hline 49 & 143.60266(52) & 143.60072(52) & 143.59895(52) & $1(0)$ & $6(1)$ & $40(8)$ \\ 81 & 328.78123(90) & 328.77826(91) & 328.77655(91) & $1(0)$ & $8(1)$ & $81(10)$ \\ 100 & 442.4766(12) & 442.4733(12) & 442.4716(12) & $1(0)$ & $9(1)$ & $110(11)$ \\ 144 & 780.8194(16) & 780.8188(16) & 780.8184(16) & $1(0)$ & $9(3)$ & $175(3)$ \\
400 & 3766.9555(39) & 3766.9515(40) & - & $1(0)$ & $19(1)$ & - \\ 
625 & 7527.5372(98) & 7527.5330(99) & - & $1(0)$ & $24(1)$ & - \\
\end{tabular} \end{ruledtabular} \end{table*}

{\it CF theory:} 
To numerically study FQH properties in the bulk, it is usually convenient to work in the spherical geometry of Haldane\cite{Haldane83}, with electrons confined to the surface of a sphere with radius $R$ and the magnetic field pointing radially outward. We note that throughout the paper, lengths are measured in units of the magnetic length $\ell_B = \sqrt{\hbar/eB}$, with $B$ being the magnetic field. The Dirac quantization condition forces the flux through the sphere to be an integer multiple of $\phi_0$, conveniently written as $2Q\phi_0$, with $Q$ being an integer or a half-integer\cite{Wu76,Wu77,Jain07}. The single-particle eigenstates are angular-momentum eigenstates, called the monopole harmonics $Y_{Qlm}(\Omega)$, where $l=|Q|+n_l$, $n_l=0,1,\dots$, and $m=-l,\dots,l$\cite{Wu76,Wu77,Haldane83}. Here $\Omega=(\theta,\phi)$ denotes the coordinates on the sphere. The quantum number $n_l$ is the Landau level (LL) index, so that $l=Q$ corresponds to the LLL. The different $m$ orbitals for a fixed $l$ are degenerate.

The CF theory posits that strongly interacting two-dimensional electrons in a perpendicular magnetic field $B$ bind an even integer number $2p$ of vortices to form weakly interacting CFs experiencing an effective magnetic field
\begin{equation}
B^*=B-2p\rho\phi_0,
\end{equation}
where $\rho$ is the density and $\phi_0$ is the flux quantum. CFs form CF LLs, or $\Lambda$ levels ($\Lambda$Ls), in the reduced magnetic field. The $\nu^*=n$ IQH effect of CFs manifests as the $\nu=n/(2pn+1)$ FQH effect of electrons, which is represented as the state of CFs completely filling the lowest $n$ \LLs. 
The many-body wave function of this state is constructed as
\begin{equation}
\psi_0=\LLL\Phi_{0}^{\nu^*=n}\Phi_1^{2p},
\end{equation}
where $\Phi_{0}^{\nu^*=n}$ is the IQH wave function at filling factor $\nu^*=n$,  the Jastrow factor $\Phi_1^{2p}=\prod_{i<j}(z_i-z_j)^{2p}$ attaches $2p$ vortices to every electron, and $\LLL$ denotes projection to the lowest Landau level. This is referred to as ``composite-fermionization." In the spherical geometry, it relates electrons at flux $2Q$ to CFs at an effective flux
\begin{equation}
2Q^*=2Q-2p(N-1).
\end{equation}
In this paper we are primarily concerned with CFs with two vortices attached, i.e. $p=1$. The CF Fermi sea corresponds to zero effective magnetic field, or $Q^*=0$. We will approach the thermodynamic limit by considering filled shell states 
with $N=n^2$~\cite{Jain07}. The LLL projection is carried out using the  quaternion formulation~\cite{Gattu25} of the Jain-Kamilla (JK) projection technique\cite{Jain97,Jain97b}. Other Slater determinant wave functions at $\nu^*$ can similarly be composite-fermionized to produce correlated wave functions at $\nu=\nu^*/(2p\nu^*+1)$. 

{\it CF diagonalization:} Consider the basis of electrons 
\begin{equation}\{\Phi_0, \{\Phi_j^{\rm KE=1}\}, \cdots \{\Phi_j^{\rm KE=\alpha}\} \}
\label{eq:basis_electron}
\end{equation}
constructed at $Q^*=0$, where $KE$ refers to the kinetic energy measured in units of the cyclotron energy relative to the ground state at $Q^*=0$. 

In the presence of translational symmetry, which is equivalent to rotational symmetry on the sphere, the many-body angular momentum operators $L^2$ and $L_z$ are good quantum numbers. The quantum number $L$ is related to momentum in the planar geometry. Exploiting the $L^2$ symmetry allows operators to be stored in block-diagonal form, reducing computational complexity.

The uniform ground state with an integral number of completely filled shells has $L=0$. States containing excitons specified by occupied orbitals $(l,m)$ are not, in general, eigenstates of $L^2$. Although the corresponding $L^2$ eigenstates may in principle be obtained by group theory, that becomes impractical when multiple excitons are present. We construct the $L^2$ eigenstates numerically by diagonalizing the many-body $L^2$ operator. Because our primary interest is in the ground state which has $L=0$, we retain only the $L=0$ block. Note that the state with single exciton with energy CFKE $=j$ has $L\geq j$ and thus is not included in our basis.

We next composite-fermionize the above electronic basis to obtain the basis 
\begin{equation}
    \{\psi_0, \{\psi_j^{\rm CFKE=1}\}, \cdots \{\psi_j^{\rm CFKE=\alpha}\} \}\label{eq:basis_CF}
\end{equation}
of correlated states at $2Q=2(N-1)$, where CFKE now is the CF kinetic energy measured in units of the CF cyclotron energy. The basis states with CFKE $=3$ are shown schematically in Fig.~\ref{fig:3ke_exciton}.

CF diagonalization (CFD) refers to diagonalizing the Coulomb interaction in this basis.   
The ground state obtained in this fashion is denoted $\psi_0^{(\alpha)}$ and its energy $E_0^{(\alpha)}$. This provides a better approximation to the true ground state than the zeroth-order approximation $\psi_0^{(0)}$.

Certain technical points deserve mention. When we composite-fermionize a collection of orthogonal states at $Q^*$ in Eq.~\ref{eq:basis_electron}, the resulting CF states in Eq.~\ref{eq:basis_CF} are, in general, neither orthogonal nor linearly independent. In fact, certain states are annihilated upon projection into the LLL. From the remaining states, we eliminate the states that have zero eigenvalues in the singular value decomposition of the overlap matrix $O_{ij}=\braket{\psi_i}{\psi_j}$. The remaining states are linearly independent but not necessarily orthogonal. The eigenvectors $c$ and eigenvalues $E$ are obtained from 
\begin{equation}
[\tilde O]^{-1}\tilde H c = E c,
\end{equation}
where the matrix $\tilde H$ is defined as $H_{ij}=\bra{\psi_i}H\ket{\psi_j}$ \cite{Gattu23,Balram15,Golub13}. The resulting eigenvalues and eigenvectors provide better approximations as higher-CFKE states are included. 

{\it Static structure factor:} We now present our results. 
The static structure factor for a state $\ket{\Psi}$ with $N$ particles is defined as \cite{He94}
\begin{equation}
S(\mathbf{q})=\frac{1}{N}\bra{\Psi}\rho_{\mathbf{q}}^{\dagger}\rho_{\mathbf{q}}\ket{\Psi}-N\delta_{\mathbf{q},0},
\end{equation}
where
\begin{equation}
\rho_{\mathbf{q}}=\sum_{j=1}^{N}e^{i\mathbf{q}\cdot\mathbf{r}_j}
\end{equation}
is the Fourier-transformed density operator, $\mathbf{r}_j$ are electron coordinates, and $\delta_{q,0}$ is the Kronecker-$\delta$ symbol. In the spherical geometry, the role of plane waves is played by spherical harmonics $Y_{LM}(\theta,\phi)$, and the momentum label $\mathbf{q}$ is replaced by angular-momentum indices $(L,M)$. Thus the relevant quantity is
\begin{equation}
S_{L,M}=\frac{4\pi}{N}\bra{\Psi}\rho_{L,M}^{\dagger}\rho_{L,M}\ket{\Psi}-N\delta_{L,0},
\end{equation}
with
\begin{equation}
\rho_{L,M}=\sum_{j=1}^{N}Y_{LM}(\theta_j,\phi_j).
\end{equation}
For a rotationally invariant state with $L=0$, the Wigner-Eckart theorem \cite{Sakurai20} implies that $S_{L,M}$ is independent of $M$.  We therefore evaluate only the $M=L$ component.

Following Ref.~\cite{Anakru25}, the mapping from the angular momentum $L$ to the wavevector $q$ is taken as $q = \sqrt{L(L+1)}/R$, which provides a better convergence to planar results at small $q$  than the often used $q = L/R$\cite{Haldane85a,Simon94a}.

When working strictly within the lowest Landau level, we use the projected static structure factor $\bar{S}(\mathbf{q})$, obtained by replacing $\rho_{\mathbf{q}}$ by its LLL-projected counterpart \cite{Girvin85,Girvin86,He94,Dora24,Kohn61}. It is related to the unprojected structure factor by
\begin{equation}
\bar{S}(\mathbf{q})=S(\mathbf{q})-[1-e^{-q^2/2}],
\end{equation}
where $q$ is measured in units of $\ell_B^{-1}$.

\begin{table}
  \caption{This table gives the overlaps $|\langle{\psi_0^{(0)}}|{\psi_0^{(2)}}\rangle|^2$, $|\langle{\psi_0^{(0)}}|{\psi_0^{(3)}}\rangle|^2$ and $|\langle{\psi_0^{(2)}}|{\psi_0^{(3)}}\rangle|^2$ for several particle numbers $N$. 
  Numbers in parentheses indicate the Monte Carlo statistical uncertainty in the last two significant digits. }
  \label{tab:ke_table_overlap}
  \centering
  \setlength{\tabcolsep}{4pt}
  \begin{ruledtabular}
  \begin{tabular}{cccc}
    $N$ & $|\langle{\psi_0^{(0)}}|{\psi_0^{(2)}}\rangle|^2$ & $|\langle{\psi_0^{(0)}}|{\psi_0^{(3)}}\rangle|^2$ & $|\langle{\psi_0^{(2)}}|{\psi_0^{(3)}}\rangle|^2$ \\
    \hline
    49  & 0.9685(10) & 0.9607(09) & 0.9860(02) \\
    64  & 0.9540(17) & 0.9478(14) & 0.9852(04) \\
    81  & 0.9415(25) & 0.9350(22) & 0.9836(05) \\
    100 & 0.9266(35) & 0.9192(33) & 0.9814(08) \\
    144 & 0.906(10) & 0.898(11) & 0.9941(08) \\
    400 & 0.836(28) & - & - \\
    625 & 0.773(79) & - & - \\
  \end{tabular}
  \end{ruledtabular}
\end{table}

We do not consider the CFKE $= 1$ excitons as they do not contain $L=0$ state. This follows because the angular momenta of the excited CF and the hole left behind are $l_{top}+1$ and $l_{top}$, respectively, which produce excitons at $l_{top}\otimes(l_{top}+1)=1\oplus2\dots\oplus(2l_{top}+1)$, where $l_{top}$ is the angular momentum of the topmost filled shell in the zeroth-order state.

\begin{figure}
    \centering
    \includegraphics[width=0.45\textwidth]{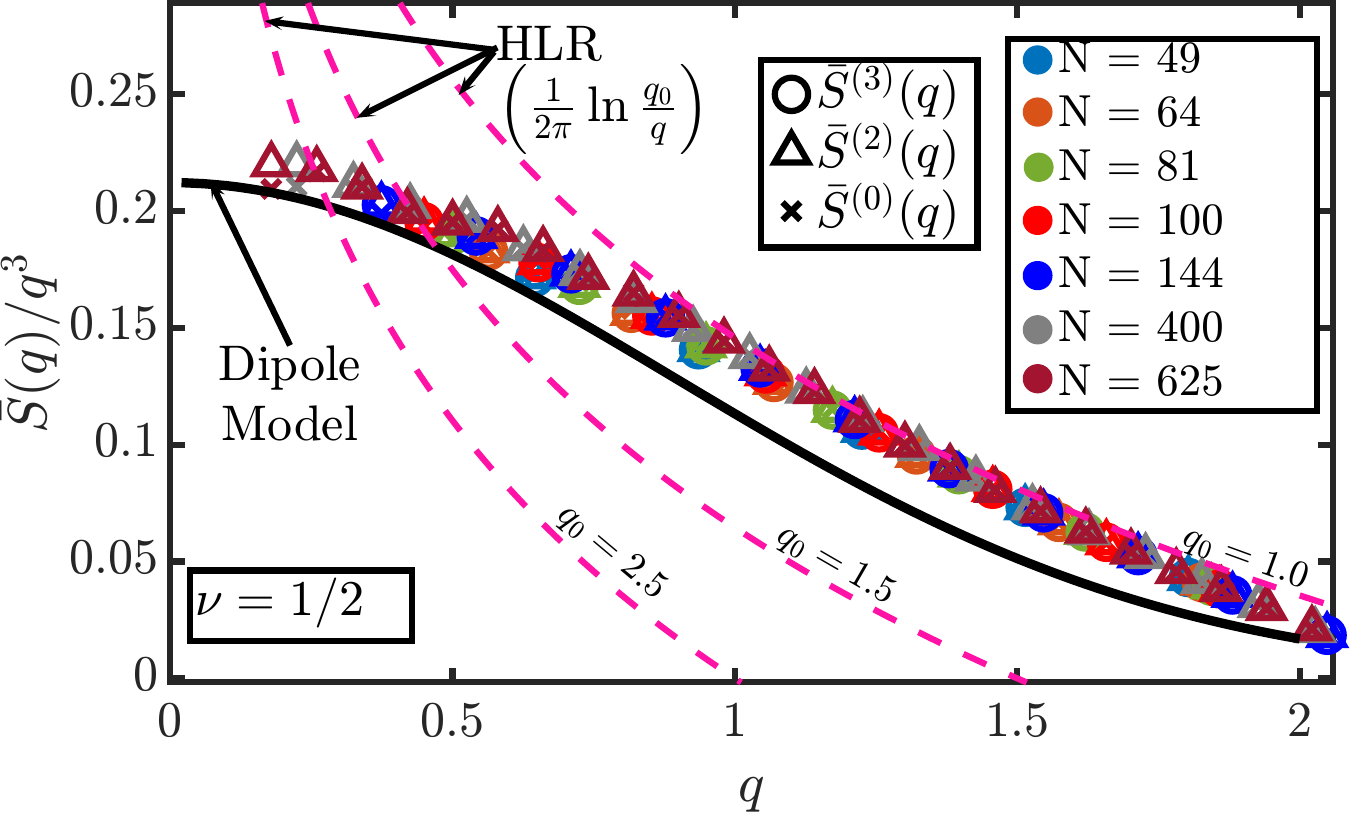}
    \caption{Plot of $\bar{S}(\mathbf{q})/q^3$ where $\bar{S}(\mathbf{q})$ is the LLL-projected static structure factor. Three different approximations for $\bar{S}(\mathbf{q})$ are shown for systems up to 625 particles. The static structure factors $\bar{S}(\mathbf{q})^{(3)}$, $\bar{S}(\mathbf{q})^{(2)}$ and $\bar{S}(\mathbf{q})^{(0)}$, calculated with $\psi_0^{(3)}$, $\psi_0^{(2)}$ and $\psi_0^{(0)}$, are shown by circles, triangles and crosses, respectively. Each color corresponds to a different system size. For systems with 400 and 625 particles we do not obtain $\bar{S}(\mathbf{q})^{(3)}$. The difference between $\bar{S}^{(3)}(\mathbf{q})$, $\bar{S}^{(2)}(\mathbf{q})$, and $\bar{S}^{(0)}(\mathbf{q})$ is not appreciable.  The magenta dashed lines show the field-theory prediction $\bar{S}(\mathbf{q}) = \frac{k_F}{2\pi}q^3\ln{\frac{q_0}{q}}$ for various values of $q_0$, and the solid black line is the prediction of the dipole model of Ref.~\cite{Anakru25}. Here $k_F = \sqrt{2\nu}$ is the CF Fermi wave vector, which equals $1/\ell_{\rm B}$ at $\nu = 1/2$. The Monte Carlo uncertainty is on the order of the symbol sizes.}
    \label{fig:Sq_over_q3}
\end{figure}

In previous works~\cite{Gattu25,Anakru25} $\psi_0^{(0)}$ had been studied for systems with as many as 900 CFs. Here, we determine $\psi_0^{(2)}$ for up to 625 CFs, and $\psi_0^{(3)}$ for up to 144 CFs. To ascertain the level of improvement, we show the ground-state energies $E_0^{\alpha}$ in Table~\ref{tab:ke_table_GSE} and overlaps in Table~\ref{tab:ke_table_overlap}.  
The overlaps $|\langle{\psi_0^{(0)}}|{\psi_0^{(3)}}\rangle|^2$ indicate significant improvement over the zeroth-order state, especially as the number of particles is increased. However, the $|\langle{\psi_0^{(2)}}|{\psi_0^{(3)}}\rangle|^2$ are greater than $0.98$ for all systems studied, indicating that the improvement in going from $\psi_0^{(2)}$ to $\psi_0^{(3)}$ is relatively small, thus strongly suggesting that $\psi_0^{(2)}$ and $\psi_0^{(3)}$ are essentially the exact CF Fermi liquid ground state. It is worth pausing for a moment to note how remarkable such a statement is for a strongly correlated state of matter.

The ground state energies also confirm this. For large systems, the dominant lowering of the energy occurs upon enlarging the basis from CFKE $=0$ to CFKE $\le 2$, while the additional energy gain obtained by going from CFKE $\le 2$ to CFKE $\le 3$ is much smaller. For $N=100$, for example, the additional lowering is within the Monte Carlo uncertainty. 

\begin{figure}[t]
    \centering
    \includegraphics[width=0.98\linewidth]{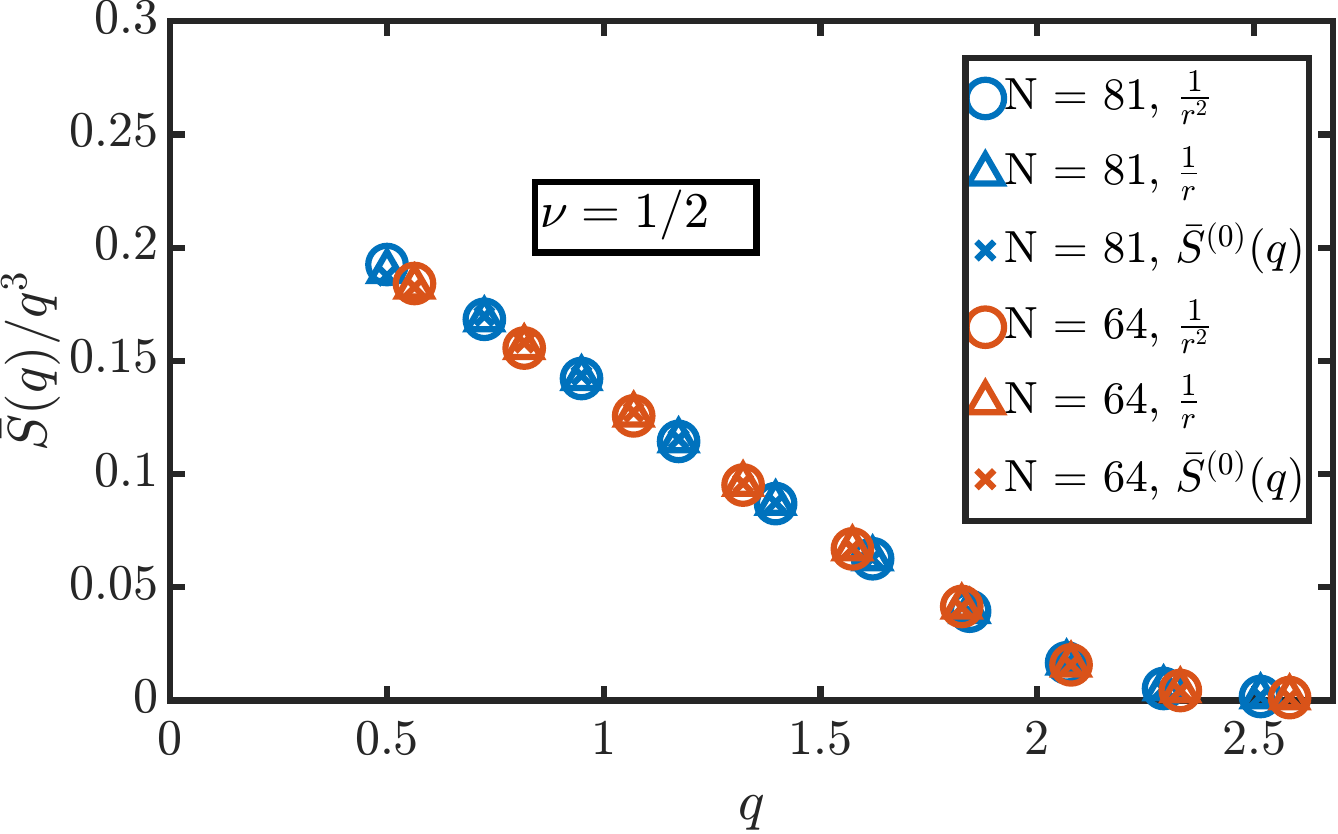}
    \caption{$\bar{S}(\mathbf{q})/q^3$ as a function of $q$ for the ground state of coulomb and $1/r^2$ interactions at 3KE level (i.e. $\bar{S}^{(3)}(\mathbf{q})$) along with the zeroth-order result. }
    \label{fig:1_r_2}
\end{figure}

We now come to the static structure factor, the main topic of this paper. Fig.~\ref{fig:Sq_over_q3} shows $\bar{S}(q)/q^3$ for various system sizes. Following Ref.~\cite{Anakru25}, we show only data corresponding to even values of $L$, which were found to converge more smoothly to the planar thermodynamic behavior. Additionally, we find empirically that the result for $L=2$ is anomalous and does not fall on the smooth curve formed by the remaining data; this anomaly is discussed in the supplementary material\cite{ssf_renorm_supl}. For all displayed points, corresponding to $L\geq4$, the zeroth-order, CFKE $\leq2$, and CFKE $\leq3$ results are nearly indistinguishable on the scale of the plot. Thus, improvement in the ground-state wave function does not produce any discernible change in the smooth long-wavelength behavior of the projected static structure factor. 
More specifically, the results from different $N$ seem to collapse nicely as a function of $q$ and show no indication of the growth expected from a $q^3\ln q$ term. Instead, $\bar{S}(q)/q^3$ flattens toward the constant value $2k_{\rm F}/3\pi$, which was earlier discussed in the context of the zeroth-order CF Fermi sea wave function and the dipole model in Ref.~\cite{Anakru25}.

{\it Interaction dependence:} To study the interaction dependence of $\bar{S}(\mathbf{q})$, we consider the interaction $V(r)\propto 1/r^2$, which corresponds to $V(q) \propto q^{-\eta}$ with $\eta=0$. 
Figure~\ref{fig:1_r_2} compares the resulting projected static structure factor $\overline{S}^{(3)}(\mathbf{q})$ for the Coulomb and $1/r^2$ interactions with the zeroth-order CF Fermi-sea result $\overline{S}^{(0)}(\mathbf{q})$.
The results for the two interactions are essentially identical.

{\it Summary:} We have obtained improved CF Fermi liquid wave functions $\psi_0^{(2)}$ and $\psi_0^{(3)}$ for up to $144$ particles, and $\psi_0^{(2)}$ for up to $625$ particles.  We find that while the CF Fermi liquid ground state improves significantly as we go from $\psi_0^{(0)}$ to $\psi_0^{(2)}$, $\psi_0^{(2)}$ and $\psi_0^{(3)}$ have $\sim 99\%$ overlap, suggesting that CFD has produced a state that is close to the exact ground state. We find that the projected static structure factor $\bar{S}(\mathbf{q})$ remains essentially the same as that obtained from the zeroth-order theory. The same remains true for a $r^{-2}$ interaction between electrons. The microscopic CF theory therefore suggests that the long wave length behavior of the static structure factor is universal, with $\bar{S}(\mathbf{q})\propto q^3$ in the limit $q\to 0$.

{\it Acknowledgments:}
We thank Prashant Kumar for insightful discussions. A. A. M. and M. G. acknowledge support in part by the National Science Foundation under Grant No. DMR-2404619. M. G. was partially supported by a 2025/2026 Rising Researcher Grant from Penn State’s Institute for
Computational and Data Sciences(RRID:~SCR\_025154). The authors of this work recognize the Penn State Institute for Computational and Data Sciences for providing access to computational research infrastructure within the Roar Core Facility (RRID:~SCR\_026424). M. G. thanks the Lodha Theoretical Physics Institute, Mumbai for its hospitality during the summer, 2026.

\bibliography{biblio_fqhe}
\newpage
\section*{Supplementary Material for Universality of long-wavelength behavior of composite-fermion Fermi liquid }

\begin{figure}
    \centering
    \includegraphics[width=0.97\linewidth]{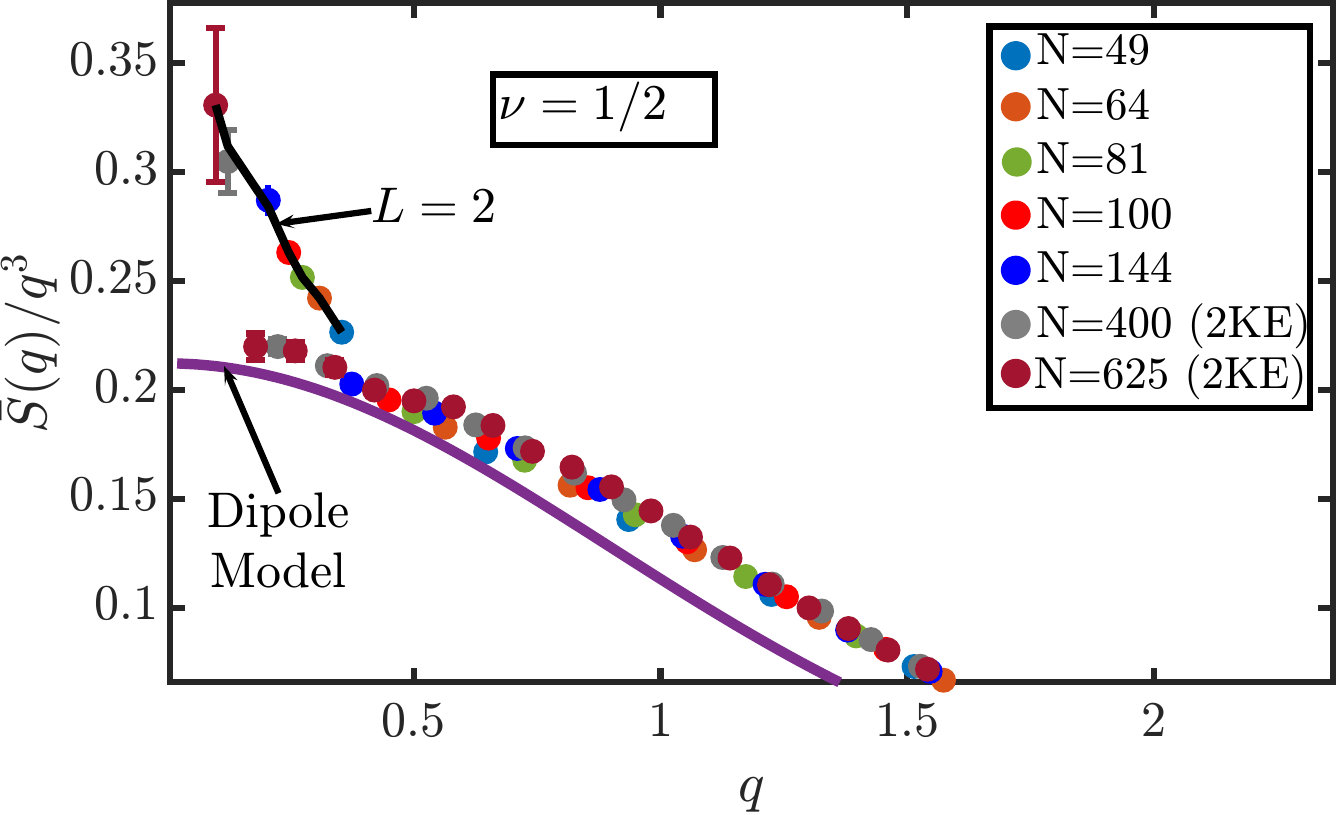}
    \caption{LLL-projected static structure factor $\bar{S}(q)/q^3$ including the $L=2$ data points omitted from Fig.~\ref{fig:Sq_over_q3}. For $N\leq144$, the results are obtained in the CFKE $\leq3$ approximation, $\bar{S}^{(3)}(q)$, whereas for $N=400$ and $625$ they are obtained in the CFKE $\leq2$ approximation, $\bar{S}^{(2)}(q)$. The black line connects the points corresponding to $L=2$ for different system sizes. These points lie away from, and do not join smoothly onto, the sequence formed by the remaining even-$L$ data with $L\geq4$, which is approximately coincident with the dipole-model result of Ref.~\cite{Anakru25}. This isolated anomalous behavior of the $L=2$ point motivates its exclusion from the analysis of the smooth thermodynamic sequence.}
    \label{fig:Sq_zoomed}
\end{figure}

\begin{figure}
    \centering
    \includegraphics[width=0.98\linewidth]{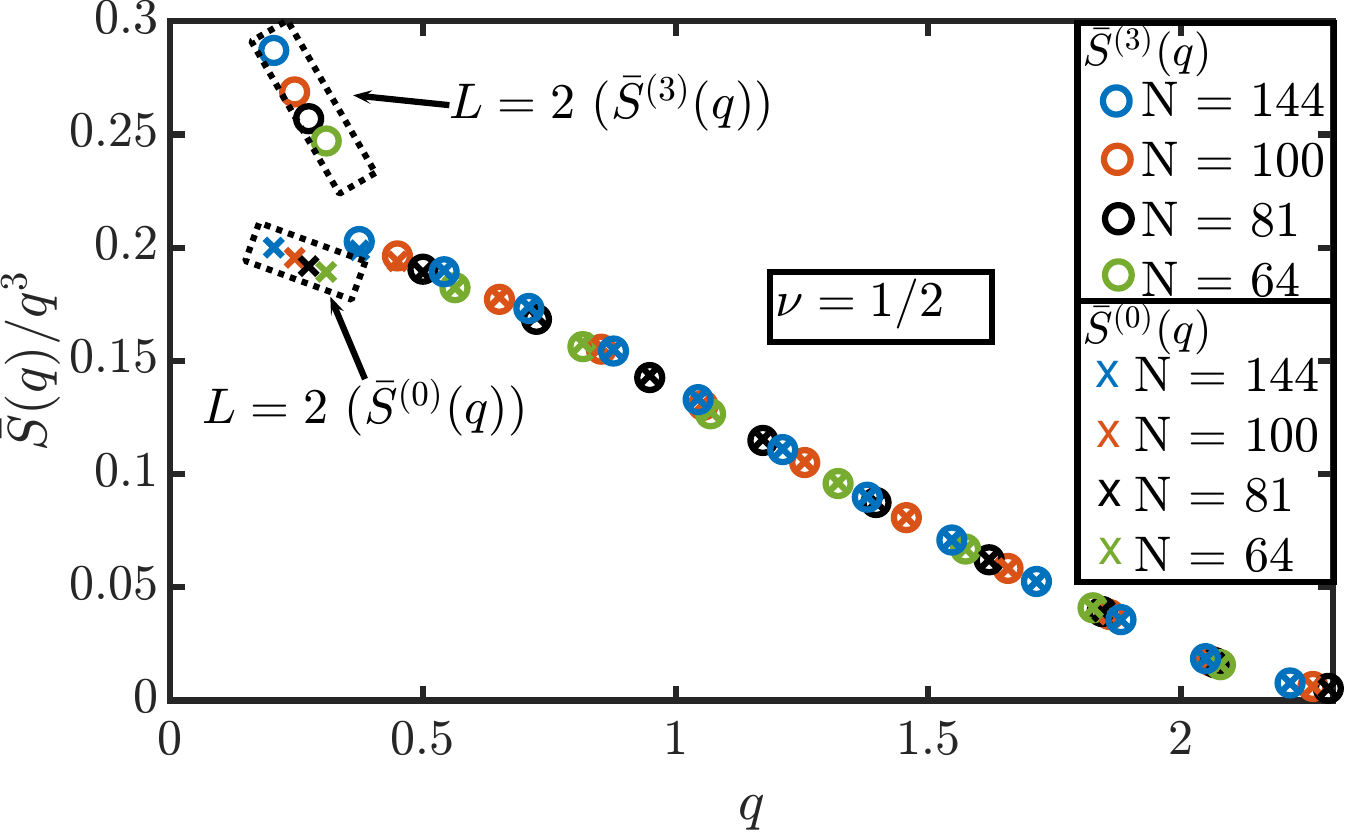}
    \includegraphics[width=0.98\linewidth]{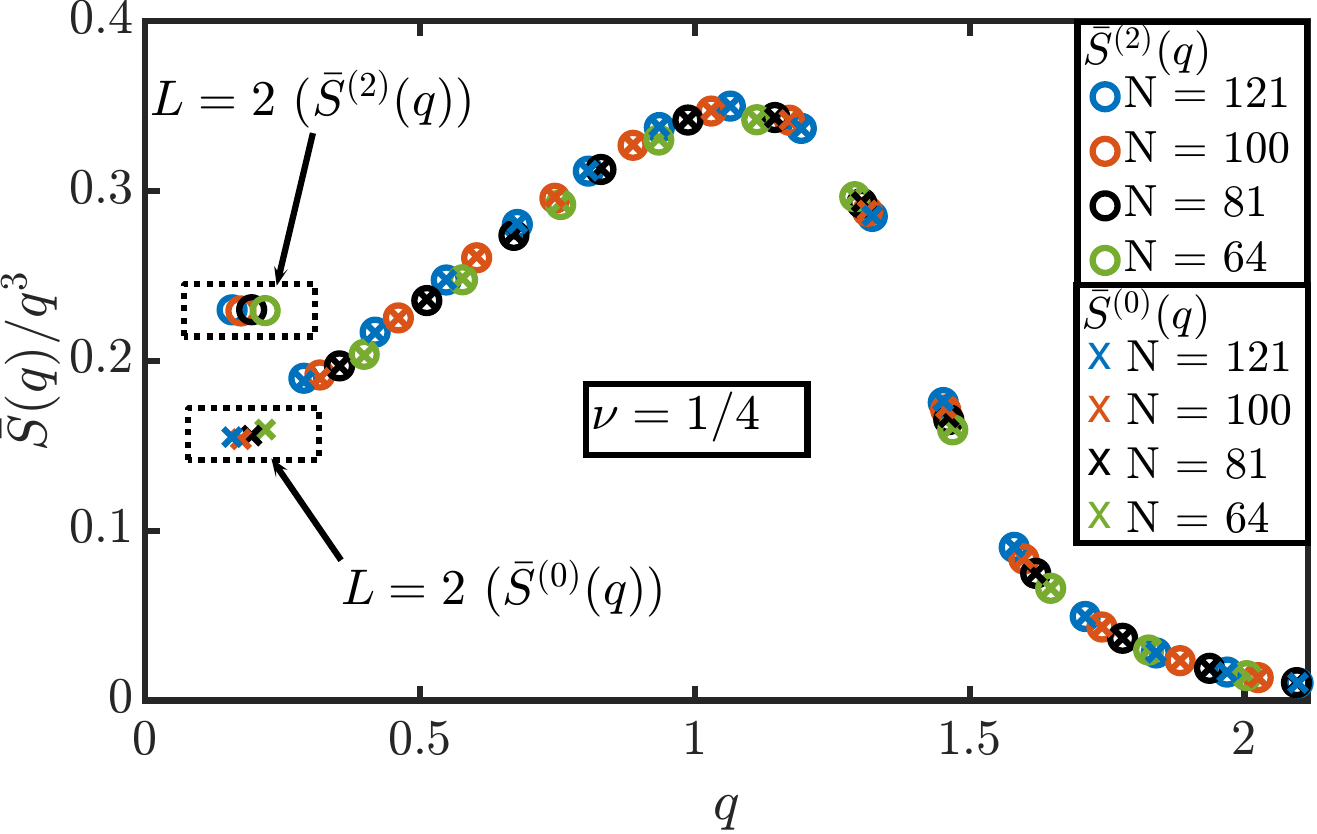}
    \caption{Comparison of the LLL-projected static structure factor at zeroth-order, $\bar{S}^{(0)}(\mathbf{q})$ (crosses) with the $\alpha$ KE corrected one, $\bar{S}^{(\alpha)}(\mathbf{q})$,(circles) at $\nu = 1/2$ (top, $\alpha = 3$) and $\nu = 1/4$ (bottom,$\alpha = 2$). The discontinuity and anomalous behavior of the $L=2$ point are more pronounced at $\nu = 1/4$ than at $\nu = 1/2$. }
    \label{fig:Sq_renorm_vs_unperturbed}
\end{figure}

In the Supplementary Material we provide empirical evidence that the static structure factor $\bar{S}(\mathbf{q})$ behaves anomalously for $L=2$, which we have therefore not considered in our thermodynamic extrapolation. 

We plot $\bar{S}(\mathbf{q})/q^3$ with $q\equiv\sqrt{L(L+1)}/R$ including the $q$ corresponding to $L=2$ in Figs.~\ref{fig:Sq_zoomed} and \ref{fig:Sq_renorm_vs_unperturbed}. Two observations are noteworthy: 

For various values of $N$, all points collapse on to a single curve with the exception of the point  corresponding to $L=2$, which lies off of this curve. In fact, this is the only point that differs appreciably from the zeroth order result, and it is visibly discontinuous from the smooth curve formed by the remaining even-$L$ data (Figs.~\ref{fig:Sq_renorm_vs_unperturbed} and \ref{fig:Sq_zoomed}). 

Furthermore, for a given $q$, the $L=2$ point from smaller $N$ systems clearly lies away from the curve obtained from $L\geq 4$ from larger $N$. For example, in Fig.\ref{fig:Sq_zoomed}, the curve for 625 particles for $L\geq 4$ clearly lies below the $L=2$ value from smaller $N$. 

Additional evidence for the anomalous behavior at $L=2$ can be seen in Fig.~\ref{fig:Sq_renorm_vs_unperturbed} which shows results for $\nu=1/4$. The $L=2$ result clearly lies off of other results.

For these reasons, we do not consider $L=2$ in our extrapolations. We do not know, however, why $S(\mathbf{q})$ behaves anomalously for $L=2$, although it is likely related to curvature effects.

\end{document}